\documentstyle[11pt]{article}

\begin{document}

\title{ON THE CHIRAL RINGS IN $N=2$ AND $N=4$ SUPERCONFORMAL
 ALGEBRAS}
\author{Murat G\"unaydin\thanks{
Work supported in part by the National Science Foundation
under Grant Number PHY-9108286.
\newline E-Mail: GXT@PSUVM or Murat@PSUPHYS1.PSU.EDU}  \\
Department of Physics \\
Penn State University \\
University Park, PA 16802 }
\date{\ \ \ }
\maketitle
\begin{abstract}
\noindent
We study the chiral primary rings of $N=2$  and $N=4$
 superconformal algebras (SCA)
constructed over triple systems.
The chiral primary states of $N=2$ SCA's realized over hermitian
Jordan triple systems are given. Their coset spaces $G/H$ are
hermitian symmetric which can be compact or non-compact. In the
non-compact case under the requirement of unitarity of the representations
of $G$ we find an infinite discrete set of chiral  primary
states associated with the holomorphic  discrete series
 representations of $G$ and their analytic continuation. Further
requirement that the corresponding $N=2$ module be unitary
 truncates this infinite set to a finite subset.
There are no chiral  primary states associated
with the other unitary representations of non-compact groups.
 Remarkably, the only non-compact  groups $G$
 that admit holomorphic discrete
 series unitary
 representations are such that their quotients $G/H$ with their
maximal compact subgroups $H$ are   hermitian symmetric.
The chiral primary states of $N=2$ SCA's constructed over the Freudenthal
triple systems are also studied. These algebras have the special property
that they admit an extension to $N=4$ superconformal algebras
with the gauge group $SU(2)\otimes SU(2) \otimes U(1)$.
We then generalize the concept of chiral rings to these maximal $N=4$
superconformal algebras. We find four different rings associated with
each sector (left or right moving). If one includes both sectors one gets
16 different rings. We also show that our analysis yields all the possible
rings of $N=4$ SCA's.

\end{abstract}

\newpage

\renewcommand{\theequation}{\arabic{section} - \arabic{equation}}
\section{Introduction}
\setcounter{equation}{0}

Extended superconformal algebras have been studied extensively in recent
years. They find fundamental applications in superstring theories
\cite{GSW} , integrable
systems \cite{IS} , topological field theories
\cite{TFT} and in the study
of critical phenomena. For example
the classical vacua of string theories are described
by conformal field theories. In particular,
the heterotic string vacua with $N = 1$ space-time
supersymmetry in four dimensions are described by  "internal" $N = 2$
superconformal field theories with  central charge $c = 9$
\cite{BDFM} which have been studied in \cite{N=2,LVW,KS89,DLP,IB}
using coset space methods \cite{GKO}.
On the other hand,
the $N = 2$ space-time supersymmetric vacua of the heterotic string
are described by an internal superconformal field theory with four
supersymmetries \cite{NS}.

The rings of chiral primary operators \cite{LVW} play a central role in the
understanding and applications of $N=2$ superconformal field
theories. In \cite{LVW} a deep connection between the cohomology rings of
Calabi-Yau manifolds and the rings of chiral primary fields of $N=2$
superconformal theories with $c=9$ was established. The existence of
 two different
rings associated with a given (2,2) superconformal theory led to the idea of
 mirror symmetry of Calabi-Yau manifolds and its natural generalization to
other complex manifolds \cite{mirror}.
Furthermore, the
chiral primary states of $N=2$ superconformal algebras turn out to be the
only physical states of topological field theories that are obtained by the
twisting of the corresponding $N=2$ superconformal theories \cite{EY}.
 In this paper we shall study the rings of chiral primary operators of
extended superconformal algebras  ($N=2$ and $N=4$). The chiral rings of $N=2$
superconformal algebras (SCA) were introduced and studied in a Cartan-Weyl
 basis
of the underlying current algebras in \cite{LVW}. Our treatment of the chiral
rings of of $N=2$ SCA's will use their realization over triple systems
\cite{MG91,GH91,GH92}. After reviewing the necessary background we study the
chiral (anti-chiral) primary operators of $N=2$ SCA's constructed over
hermitian Jordan triple systems which correspond to their realization over
hermitian
symmetric spaces $G/H$
 that may be compact or non-compact. We give a complete characterization of the
chiral primary states both in the compact and the non-compact
cases. In the non-compact case under the requirement of unitarity of
the representations of $G$ at every level we find an infinite discrete set
of chiral (anti-chiral) primary states associated with the holomorphic
(anti-holomorphic) discrete series representations of $G$ and their
analytic continuation. However, the requirement of unitarity of the $N=2$
module truncates this infinite discrete set to a finite subset. The unitary
representations of the noncompact groups outside the holomorphic (or anti-
holomorphic ) discrete series representations
and their analytic continuation do not lead to any non-trivial
chiral (or anti-chiral) primary states. Remarkably , the only non-compact
groups $G$ that admit holomorphic (anti-holomorphic) discrete series
representations are those whose quotient $G/H$ with their maximal compact
subgroups $H$ are hermitian symmetric. A finite set of purely ``fermionic''
chiral primary states exist in both the compact and the non-compact cases.
Next we study the chiral primary states in $N=2$ SCA's constructed over
Freudenthal triple systems (FTS). The $N=2$ SCA's constructed over FTS's have
the very
special property that they allow an extension to $N=4$ SCA's with the gauge
group $SU(2) \otimes SU(2) \otimes U(1)$ \cite{GH91,GH92}. In the second part
of the paper we generalize the concept of chiral (anti-chiral) rings to
these maximal $N=4$ SCA's. We find that the natural extension of the concept
of a chiral ring to $N=4$ SCA's leads to four different rings associated with
each sector (left or right moving). If we include both sectors we find
sixteen different rings. We also show that our analysis yields all the
 possible
rings of $N=4$ SCA's. Throughout the paper we use the Neveu-Schwarz moding.
However, our results can easily be carried over to the Ramond moding using
spectral flow.

\section{Construction of $N=2$ Superconformal Algebras over Jordan Triple
 Systems}

\setcounter{equation}{0}
In this section we shall review the construction of $N=2$ superconformal over
Jordan triple systems \cite{MG91} and reformulate it in such a way as to make
it easier to relate to the coset space methods that use the Cartan-Weyl basis
for Lie algebras.
Consider a 3-graded Lie algebra $g$:
\begin{equation}
g = g^{-1} \oplus g^{0} \oplus g^{+1}
\end{equation}
where $ \oplus $ denotes vector space direct sum and $g^{0}$ is a
subalgebra of maximal rank. We have the formal commutation relations of the
 elements of various grade subspaces
\begin{equation}
[g^{m},g^{n}] \subseteq g^{m+n} \; ; m,n=-1,0,1
\end{equation}
where $ g^{m+n} = 0 $ if $|m+n|>1$. Every simple Lie algebra with such a
3-graded structure can be constructed in terms of an underlying
Jordan triple system (JTS) $V$ via the Tits-Kantor-Koecher (TKK)
  construction \cite{TKK}. This construction establishes a one-to-one mapping
between the grade $+1$ subspace of $g$ and the underlying JTS $V$:
\begin{equation}
U_{a} \in g  \Longleftrightarrow  a \in V
\end{equation}
Every such Lie algebra $g$ admits
 a conjugation (involutive automorphism) $\dagger$
under which the elements of the grade $+1$ subspace get mapped into the
elements of the grade $-1$ subspace.
\begin{equation}
U^{a} = U_{a}^{\dagger}  \in g^{-1}
\end{equation}
One then defines
\begin{equation}
\begin{array}{l}
[U_{a},U^{b}] = S_{a}^{b} \\
\  \\
{[}S_{a}^{b}, U_{c}{]} = U_{(abc)}
\end{array}
\end{equation}
where $S_{a}^{b} \in g^{0}$ and $(abc)$ is a triple product under which the
 elements of $V$ close. Under conjugation $\dagger$ one finds
\begin{equation}
\begin{array}{l}
(S_{a}^{b})^{\dagger} = S_{b}^{a} \\
\  \\
{[}S_{a}^{b},U^{c}{]} = - U^{(bac)}
\end{array}
\end{equation}
The Jacobi identities in $g$ are satisfied if and only if the ternary
product $(abc)$ satisfies the defining identities of a JTS:
\begin{equation}
\begin{array}{l}
(abc) = (cba)  \\
\  \\
(ab(cdx))-(cd(abx))-(a(dcb)x)+((cda)bx) = 0
\end{array}
\end{equation}
The elements $S_{a}^{b}$ of the grade zero subspace form a subalgebra
which is called the structure algebra of $V$ :
\begin{equation}
[S_{a}^{b},S_{c}^{d}] = S_{(abc)}^{d}-S_{c}^{(bad)}=S_{a}^{(dcb)}
-S_{(cda)}^{b}
\end{equation}

 Consider now the affine Lie algebra (current algebra) $\hat{g}$ defined by
$g$. The commutation relations of  $\hat{g}$ can be written as
operator products in the basis defined by the underlying JTS:
\begin{equation}
\begin{array}{lll}
U_{a}(z) U^{b}(w)& =& \frac{k \delta_{a}^{b}}{(z-w)^{2}}
+ \frac{1}{(z-w)} S_{a}^{b}(w) + \cdots  \\
\  & \ & \  \\
S_{a}^{b}(z) U_{c}(w)& =& \frac{1}{(z-w)} U_{(abc)} + \cdots \\
\ & \ & \  \\
S_{a}^{b}(z) U^{c}(w) & = & \frac{-1}{(z-w)} U^{(bac)}+ \cdots  \\
\ & \ & \  \\
S_{a}^{b}(z) S_{c}^{d}(w) & = & \frac{k \Sigma_{ac}^{bd}}{(z-w)^{2}}
+ \frac{1}{(z-w)} \{ S_{(abc)}^{d} - S_{c}^{(bad)} \}(w) + \cdots
\end{array}
\end{equation}
where k is the level of $\hat{g}$ and $\Sigma_{ac}^{bd}$ are the structure
 constants of the JTS $V$:
\begin{equation}
U_{(abc)} = \Sigma_{ac}^{bd} U_{d}
\end{equation}
We choose a basis for the JTS such that the structure constants satisfy

\begin{equation}
\begin{array}{l}
\Sigma_{ae}^{bf} \Sigma_{fc}^{ed} = \check{g} \Sigma_{ac}^{bd}  \\
\  \\
\Sigma_{ac}^{bc} = \check{g} \delta_{a}^{b}
\end{array}
\end{equation}
where $\check{g}$ is the dual coxeter number of the Lie algebra $g$.
We introduce complex Fermi fields labelled by the elements of $V$ that
 satisfy
\begin{equation}
\begin{array}{l}
\psi_{a}(z) \psi^{b}(w) = \frac{\delta_{a}^{b}}{(z-w)} + \cdots    \\
\  \\
\psi_{a}(z) \psi_{b}(w) = \psi^{a}(z) \psi^{b}(w) = 0 + \cdots
\end{array}
\end{equation}
The supersymmetry generators defined by the following  bilinears of
the fermions and the currents labelled by the elements of $V$
\begin{equation}
\begin{array}{l}
G(z) = \sqrt{\frac{2}{(k+\check{g})}} U_{a} \psi^{a}(z)  \\
\  \\
\bar{G}(z) = \sqrt{\frac{2}{(k+\check{g})}} U^{a} \psi_{a} (z)
\end{array}
\end{equation}
generate an $N=2$ SCA
\footnote{All local composite operators with a single argument are
assumed to be normal ordered. Our conventions for normal ordering are
the same as those of reference \cite{BBSS}.}

\begin{equation}
G(z) \bar{G}(w) = \frac{\frac{2c}{3}}{(z-w)^{3}} +
\frac{2J(w)}{(z-w)^{2}} + \frac{2T(w)+ \partial J(w)}{(z-w)} + \cdots
\end{equation}
with central charge
\begin{equation}
c = \frac{3kD}{(k+\check{g})}
\end{equation}
where $D$ is the dimension of the JTS $V$. The Virasoro generator
$T(z)$ and the U(1) current $J(z)$ of the $N=2$ SCA are given by
\begin{equation}
\begin{array}{l}
T(z) = \frac{1}{(k+\check{g})} \{ \frac{1}{2} (U^{a} U_{a} + U_{a}
U^{a}) - \frac{k}{2} (\psi_{a} \partial \psi^{a} + \psi^{a} \partial
\psi_{a}) + S_{a}^{b} \psi^{a} \psi_{b} \}(z)  \\
\  \\
J(z) = \frac{1}{(k+\check{g})} \{ S_{a}^{a} + k \psi^{a} \psi_{a} \}(z)
\end{array}
\end{equation}
Note that $S_{a}^{a}$ is the trace of the structure algebra of $V$.

If we denote the groups generated by the Lie algebra $g$ and its subalgebra
$g^{0}$ as $G$ and $H$, respectively, then it can be shown that the
above realization of $N=2$ SCA's over JTS's is equivalent to their construction
over the symmetric spaces $G/H$ \cite{MG91}.
A JTS is called hermitian if it can be given the structure of a
complex vector space such that the triple product (abc) is linear in
the first and the last arguments and anti-linear in the second
argument. The coset spaces $G/H$ corresponding to  hermitian
JTS's are hermitian symmetric spaces. The complete list of simple hermitian
JTS's includes four infinite families and two exceptional ones
\cite{OL}. In the above we implicitly assumed that we are working with
JTS's underlying compact Lie groups.
To obtain the realizations of $N=2$ SCA's corresponding to non-compact
symmetric spaces $G/H$ where $H$ is the maximal compact subgroup of $G$
one needs simply to replace the ternary product
$(abc)$ of the compact case by its negative $-(abc)$.
This introduces an overall minus sign in the signatures of
the Killing metric associated with the coset space
generators. The central term of the current algebra $\hat{g}$
must then be taken as $-k$ times the Killing metric in order for the
contribution to the central charge from the non-compact generators to
 be  positive for positive $k$. This is then equivalent
to replacing the dual coxeter number $\check{g}$ by its negative
in the expression for the central charge. Therefore the central
charge for the non-compact realization of $N=2$ SCA's is \cite{DLP,IB,MG91}
\begin{equation}
c = \frac{3kD}{(k - \check{g})}
\end{equation}
The realization of $N=2$ superconformal algebras over non-compact coset
spaces were studied in \cite{DLP,IB} and the cohomology of non-compact
coset models in \cite{LZ}.
Let us now reformulate the above construction so as to relate it to the
one given by Kazama and Suzuki \cite{KS89}. We first note that the
the quadratic Casimir operator of the compact
 Lie algebra contructed over a JTS, in our normalization, is given by
\begin{equation}
C_{2} =  \frac{1}{2} ( U_{a} U^{a} + U^{a} U_{a} ) + \frac{1}{2 \check{g}}
S_{a}^{b}S_{b}^{a}
\end{equation}
 Again to obtain the Casimir operator of the non-compact Lie algebra
with maximal compact subalgebra $g^{0}$ one needs simply to replace the
dual Coxeter number $\check{g}$ by its negative.
The following fermion bilinears
\begin{equation}
\bar{S}_{a}^{b} = - \Sigma_{ac}^{bd} \psi^{c} \psi_{d} (z)
\end{equation}
generate the current algebra  isomorphic to $\hat{g}^{0}$
\begin{equation}
\bar{S}_{a}^{b}(z) \bar{S}_{c}^{d}(w) =
\frac{\check{g} \Sigma_{ac}^{bd}}{(z-w)^{2}} +
\frac{\bar{S}^{d}_{(abc)}(w)}{(z-w)} -
\frac{\bar{S}^{(bad)}_{c}(w)}{(z-w)} + ...
\end{equation}
By the symmetric space theorem of \cite{GNO} one can rewrite the bilinears
in the currents $\bar{S}_{a}^{b}$ as bilinears in the fermions
\begin{equation}
\frac{1}{\check{g}} \bar{S}_{a}^{b} \bar{S}_{b}^{a}(w)=
- \check{g} ( \psi^{a} \partial \psi_{a} + \psi_{a} \partial \psi^{a})(w)
\end{equation}
where the factor $\frac{1}{\check{g}}$ on the left hand side follows from
our normalization of the generators. Using these results we can write the
term $ S_{a}^{b} \psi^{a} \psi_{b}(z) $ that appears in the expression
for $T(z)$ as
\begin{equation}
\begin{array}{lll}
S_{a}^{b} \psi^{a} \psi_{b}(z) & = & - \frac{1}{2\check{g}}
(S_{a}^{b} + \bar{S}_{a}^{b} )( S_{b}^{a} + \bar{S}_{b}^{a})(z) \\
\ & \ &  \   \\
\ & \ & + \frac{1}{2\check{g}} S_{a}^{b} S_{b}^{a}(z)
- \frac{\check{g}}{2} ( \psi^{a} \partial \psi_{a} + \psi_{a} \partial \psi^{a}
)(z)
\end{array}
\end{equation}
where we used the identity
\begin{equation}
\Sigma_{ac}^{bd} S_{d}^{c} = \check{g} S_{a}^{b}
\end{equation}
Substituting this in the expression for $T(z)$ we obtain
\begin{equation}
\begin{array}{lll}
T(z) &=& \frac{1}{(k+\check{g})} \{ \frac{1}{2}(U_{a}U^{a} + U^{a}U_{a})
+ \frac{1}{2\check{g}} S_{a}^{b}S_{b}^{a} \}(z)  \\
\ & \ & \  \\
\ & \ &  - \frac{1}{(k+\check{g})} \frac{1}{(2\check{g})}
\{( S_{a}^{b}+\bar{S}_{a}^{b})(S_{b}^{a}+\bar{S}_{b}^{a}) \}(z) \\
\ & \ & \  \\
\ & \ & - \frac{1}{2} \{ \psi_{a} \partial \psi^{a} + \psi^{a}
\partial \psi_{a} \}(z)
\end{array}
\end{equation}
which shows that it is precisely the Virasoro generator associated with the
coset space
\begin{equation}
\frac{G_{k} \times SO_{1}(2D)}{H_{k+g-h}}
\end{equation}
,thus agreeing with Kazama and Suzuki \cite{KS89}.
\\
\section{Chiral Primary States of $N=2$ Superconformal Algebras realized over
 Hermitian Jordan Triple Systems}
\setcounter{equation}{0}
In the Neveu-Schwarz moding the generators of $N=2$ SCA's realized over the
Hermitian Jordan triple systems are given by
\begin{equation}
\begin{array}{lll}
G_{s} & = & \sqrt{\frac{2}{k+\check{g}}} \sum_{n} U_{an} \psi^{a}_{s-n} \\
\ & \ & \ \\
\bar{G}_{s} & = & \sqrt{\frac{2}{k+\check{g}}} \sum_{n} U_{n}^{a}
\psi_{a(s-n)}  \\
\ & \ & \  \\
J_{n} & = & \frac{1}{k+\check{g}} \{ S_{an}^{a} + k \sum_{s}
: \psi^{a}_{n-s} \psi_{as}: \}  \\
\ & \ & \  \\
T_{n} & = & \frac{1}{k+\check{g}} \{ \frac{1}{2} \sum_{m} :( U^{a}_{(n-m)}
U_{am} + U_{a(n-m)} U^{a}_{m} ): \\
\ & \ & \  \\
\ & \ & + \frac{k}{2} \sum_{s} (s+\frac{1}{2}):( \psi_{a(n-s)} \psi^{a}_{s}
+ \psi^{a}_{(n-s)} \psi_{as} ): + \sum_{m,s} S_{a(n-m)}^{b}
:\psi^{a}_{(m-s)} \psi_{bs}: \}  \\
\ & \ & \  \\
\ & \ & m,n,... = 0, \pm1, \pm2,... \\
\ & \ & \  \\
\ & \ & r,s,... = \pm\frac{1}{2}, \pm\frac{3}{2}, ...
\end{array}
\end{equation}
The primary states $ |h,q \rangle $ of the $N=2$ SCA are defined
by the conditions
\begin{equation}
\begin{array}{l}
T_{0} |h,q \rangle = h |h,q \rangle \\
\  \\
J_{0} |h,q \rangle = q |h,q \rangle  \\
\  \\
T_{n} |h,q \rangle = J_{n} |h,q \rangle =0 , \; n>0 \\
\  \\
G_{r} |h,q \rangle = \bar{G}_{r} |h,q \rangle =0 , \; r>0
\end{array}
\end{equation}
The chiral primary states satisfy the additional condition
\begin{equation}
G_{-\frac{1}{2}} | h,q \rangle = 0
\end{equation}
which implies that \cite{LVW}
\begin{equation}
 h=\frac{q}{2}
\end{equation}
On the other hand, the anti-chiral primary
states satisfy
\begin{equation}
\bar{G}_{-\frac{1}{2}}| h,q \rangle = 0  \Leftrightarrow  h = -\frac{q}{2}
\end{equation}
In order to identify the chiral and anti-chiral primary states let us
write the operators $T_{0},J_{0},G_{\pm \frac{1}{2}},
 \bar{G}_{\pm \frac{1}{2}}$ in a more explicit form
\begin{equation}
\begin{array}{lll}
T_{0} & = & \frac{1}{(k+\check{g})} \{ \frac{1}{2} \sum_{m} :( U^{a}_{-m}
U_{am} + U_{a(-m)} U^{a}_{m} ):  \\
\ & \ & \  \\
\ & \ & +k \sum_{s>0} s:( \psi^{a}_{-s} \psi_{as} + \psi_{a(-s)} \psi^{a}_{s}
): + \sum_{m,s} S^{b}_{a(-m)}: \psi^{a}_{(m-s)}\psi_{bs}: \} \\
\ & \ & \ \\
J_{0} & = & \frac{1}{k+\check{g}} \{ S^{a}_{a0} + k \sum_{s>0} s:(
\psi^{a}_{-s} \psi_{as} - \psi_{a(-s)} \psi^{a}_{s} ): \} \\
\ & \ & \  \\
G_{\pm \frac{1}{2}}& = & \sqrt{\frac{2}{(k+\check{g})}} \{ U_{a0}
\psi^{a}_{\pm \frac{1}{2}} + \sum_{n>0} ( U_{an} \psi^{a}_{\pm \frac{1}{2} -n}
+ U_{a(-n)} \psi^{a}_{\pm \frac{1}{2}+n} ) \} \\
\ & \ & \  \\
\bar{G}_{\pm \frac{1}{2}} & = & \sqrt{\frac{2}{(k+\check{g})}} \{ U^{a}_{0}
\psi_{a(\pm \frac{1}{2})} + \sum_{n>0} ( U^{a}_{n} \psi_{a(\pm \frac{1}{2}-n)}
 + U^{a}_{-n} \psi_{a(\pm \frac{1}{2}+n)} ) \}
\end{array}
\end{equation}
As is standard in the literature we shall use the term ``vacuum representation
of $\hat{g}$'' to refer to the states belonging to a highest
weight representation
 that are annihilated by all the positive mode generators of $\hat
{g}$ and that form a representation of $g$.
The components $|\Omega \rangle$
 of the vacuum representation of $\hat{g}$
that are annihilated by the grade $+1$ operators $U_{a0}$  are
chiral primary since \footnote{Note that the generators of the $N=2$
superconformal algebra act on the tensor product of the Fock space of the
fermions and the representation space of the affine Lie algebra $\hat{g}$.
Therefore the state $|\Omega \rangle $ represents the tensor product of the
Fock vacuum with the state $\Omega$ of the vacuum representation
of $\hat{g}$.}
\begin{equation}
\begin{array}{l}
T_{0} |\Omega \rangle = \frac{1}{2(k+\check{g})} S_{a0}^{a} | \Omega \rangle
= \frac{p}{2(k+\check{g})} | \Omega \rangle  \\
\  \\
J_{0} | \Omega \rangle = \frac{1}{(k+\check{g})} S_{a0}^{a} | \Omega \rangle
= \frac{p}{(k+\check{g})} | \Omega \rangle
\end{array}
\end{equation}
where $p$ is the eigenvalue of the trace current $S_{a0}^{a}$ of the
structure algebra on the state $|\Omega \rangle $.
Let $|0\rangle $ denote the tensor product of the Fock vacuum with the
scalar vacuum representation of $\hat{g}$. It satisfies
\begin{equation}
\begin{array}{l}
\psi_{r}^{a}|0 \rangle = \psi_{ar} |0 \rangle = 0 , \; r>0   \\
\  \\
U_{an} |0\rangle = U^{a}_{n}|0\rangle = 0  \; , n\geq 0
\end{array}
\end{equation}
Then the states of the form
\begin{equation}
|a_{1}, ..., a_{N} \rangle \equiv  \psi^{a_{1}}_{-\frac{1}{2}}
 \cdots \psi^{a_{N}}_{-\frac{1}{2}} | 0 \rangle
\end{equation}
 are all chiral primary.
The eigenvalues of $T_{0}$ and $J_{0}$ are
\begin{equation}
\begin{array}{l}
T_{0}|a_{1},...,a_{N} \rangle = \frac{kN}{2(k+\check{g})}
 |a_{1},...,a_{N} \rangle \\
\  \\
J_{0} |a_{1},...,a_{N} \rangle = \frac{kN}{(k+\check{g})}
|a_{1},...,a_{N} \rangle
\end{array}
\end{equation}
As for the anti-chiral primary states they are of the form
\begin{equation}
|\bar{a}_{1},...,\bar{a}_{N} \rangle \equiv \psi_{a_{1}(-\frac{1}{2})} \cdots
\psi_{a_{N}(-\frac{1}{2})} |0 \rangle
\end{equation}
or of the form $ | \bar{\Omega} \rangle $ which are states belonging to the
vacuum representation of $\hat{g}$ annihilated by
the grade $-1$ generators $U^{a0}$
\begin{equation}
U^{a0} |\bar{\Omega} \rangle = 0
\end{equation}
For the anti-chiral primary states we find
\begin{equation}
\begin{array}{l}
T_{0} | \bar{\Omega} \rangle = \frac{-1}{2(k+\check{g})} S_{a0}^{a}
|\bar{\Omega} \rangle = \frac{\bar{p}}
{2(k+\check{g})} |\bar{\Omega} \rangle \\
\ \\
J_{0} | \bar{\Omega} \rangle = \frac{-\bar{p}}{2(k+\check{g})}
|\bar{\Omega} \rangle  \\
\ \\
T_{0} |\bar{a}_{1},...,\bar{a}_{N} \rangle = \frac{kN}{2(k+\check{g})}
|\bar{a}_{1},...,\bar{a}_{N} \rangle \\
\  \\
J_{0} |\bar{a}_{1},...,\bar{a}_{N} \rangle = \frac{-kN}{(k+\check{g})}
|\bar{a}_{1},...,\bar{a}_{N} \rangle
\end{array}
\end{equation}
Note that the eigenvalues of the trace operator $S_{a0}^{a}$ on the states
 $|\Omega \rangle $ and $|\bar{\Omega} \rangle $ are non-negative and
non-positive, respectively.

The unique chiral and anti-chiral primary states
 with $h=\frac{c}{6}$ are obtained by the
action of $D$ copies of the fermion operators $\psi_{-\frac{1}{2}}^{a}$
and $ \psi_{a(-\frac{1}{2})}$ on the Fock vacuum
\begin{equation}
\begin{array}{l}
|h=\frac{kD}{2(k+\check{g})}, q=\frac{kD}{(k+\check{g})} \rangle =
|a_{1},....,a_{D} \rangle  \\
\  \\
|h=\frac{kD}{2(k+\check{g})}, q=-\frac{kD}{(k+\check{g})} \rangle =
|\bar{a}_{1},..., \bar{a}_{D} \rangle
\end{array}
\end{equation}
and are annihilated by $G_{-\frac{3}{2}}$ and $\bar{G}_{-\frac{3}{2}}$,
respectively.
 The additional chiral primary states are obtained by tensoring the
anti-chiral primary states $|\bar{\Omega} \rangle $ with the unique
chiral primary state $|a_{1},...,a_{D} \rangle $
\begin{equation}
|\bar{\Omega} , a_{1},...,a_{D} \rangle \equiv
|\bar{\Omega} \rangle \otimes |a_{1},...,a_{D} \rangle
\end{equation}
They satisfy
\begin{equation}
\begin{array}{lll}
T_{0}|\bar{\Omega},a_{1},...a_{D} \rangle & = &
\frac{1}{2(k+\check{g})} \{ S^{a}_{a0} +kD \} |\bar{\Omega},a_{1},...a_{D}
\rangle = \frac{(kD-\bar{p})}{2(k+\check{g})} |\bar{\Omega},a_{1},...,a_{D}
\rangle  \\
\ & \ & \  \\
J_{0} |\bar{\Omega},a_{1},...,a_{D} \rangle & = &
\frac{1}{(k+\check{g})} \{ S^{a}_{a0} + kD \}|\bar{\Omega},a_{1},...,a_{D}
\rangle  = \frac{(kD-\bar{p})}{(k+\check{g})}
 |\bar{\Omega},a_{1},...,a_{D} \rangle
\end{array}
\end{equation}
where we used the fact that
\begin{equation}
\psi^{a}_{-\frac{1}{2}} \psi_{b(\frac{1}{2})} |a_{1},...,a_{D} \rangle
= \delta^{a}_{b} |a_{1},...,a_{D} \rangle
\end{equation}
The corresponding anti-chiral primary states are obtained by tensoring
the states $|\Omega \rangle $ with the unique anti-chiral primary state
$|\bar{a}_{1},...,\bar{a}_{D} \rangle $ which satisfy
\begin{equation}
\begin{array}{l}
T_{0}|\Omega,\bar{a}_{1},..,\bar{a}_{D} \rangle =
\frac{(kD-p)}{2(k+\check{g})} |\Omega,\bar{a}_{1},...,\bar{a}_{D} \rangle \\
\  \\
J_{0} |\Omega, \bar{a}_{1},...,\bar{a}_{D} \rangle =
\frac{(p-kD)}{(k+\check{g})} |\Omega,\bar{a}_{1},...,\bar{a}_{D} \rangle
\end{array}
\end{equation}

Let us now try to extend the above results to the case when the affine Lie
algebra $\hat{g}$ is non-compact. It is known that the non-compact affine
 Lie algebras
do not admit any non-trivial unitary representations of the highest weight
type. The only
known method for constructing unitary theories over them appears to be the
coset space method \cite{GKO,DLP,IB,LZ}.
 For non-compact groups $G$ the relevant coset
 space is $G/H$ where $H$ is the maximal compact subgroup of $G$. However, all
such coset spaces are symmetric spaces. Furthermore, the requirement that the
unitary realization of the Virasoro algebra admit extension to $N=2$
supersymmetry implies that $G/H$ must be  Kaehlerian . The irreducible
non-compact Kaehlerian symmetric spaces are precisely the non-compact
Hermitian symmetric spaces \cite{KN}.  Therefore in the non-compact case
we need only to restrict ourselves to the $N=2$ SCA's realized over  the
Hermitian Jordan triple systems. The formalism we developed is very well
suited for studying the non-compact case in complete analogy with the compact
 case. Of course there are many novelties and subtleties that have to be taken
into account in the non-compact case. We shall start our discussion by
 assuming that the only representations of $G$ of physical relevance are
 unitary representations. This is a necessary condition for a unitary
 realization of the superconformal algebra. However, it is by no means
sufficient. In fact , it is not even sufficient for unitarity of the
representations
of the affine Lie algebra $\hat{g}$.  Now for non-compact $G$ the
generators $U_{a}$ and $U^{a}$ of grade $+1$ and $-1$ are the non-compact
generators. As was apparent in the discussion of the compact case a large class
 of chiral (anti-chiral) primary states were  defined by those states  in the
vacuum representation of $\hat{g}$ that are annihilated by the
grade $+1$ ($-1$) operators $U_{a0} (U^{a0})$. By our assumption the
vacuum representation of $\hat{g}$ is a unitary representation of $g$.
However, in general  a unitary representation of a non-compact Lie algebra
 $g$ does not contain any states that are annihilated by all the $U_{a0}$ or
all the $U^{a0}$. This happens precisely when the Lie algebra $g$
(or rather the non-compact group G corresponding to it) admits
holomorphic or anti-holomorphic unitary representations belonging to the
discrete series or its analytic continuation \cite{HC}.
 Remarkably , the
necessary and sufficient condition for the existence of such representations
is that the quotient space $G/H$ be hermitian symmetric, where $H$ is the
maximal compact subgroup of $G$.  For each such $G$
there exists a discrete infinity  of holomorphic and anti-holomorphic
unitary representations. They can be realized over the Fock spaces of
a set of bosons transforming in a certain representation of its maximal
compact subgroup $H$ and have been studied extensively over the last
decade in the physics literature \cite{oscmet}.

As in the compact case let us denote by $|\Omega \rangle$ and $|\bar{\Omega}
\rangle$ the states belonging to the vacuum representation that are
 annihilated by $U_{a0}$ and $U^{a0}$, respectively. Then
the chiral primary states are of three types
\begin{equation}
\begin{array}{l}
i) \hspace{1.0cm} |\Omega \rangle \\
\ \\
ii) \hspace{1.0cm} |a_{1},...a_{N} \rangle \hspace{1.0cm} ,  N=1,...,D \\
\  \\
iii) \hspace{1.0cm} |\bar{\Omega},a_{1},...,a_{D}\rangle
\end{array}
\end{equation}

The eigenvalues of $T_{0}$ and $J_{0}$ on these states are

\begin{equation}
\begin{array}{l}

T_{0} | \Omega \rangle = \frac{1}{2(k-\check{g})} S^{a}_{a0} | \Omega
\rangle = \frac{p}{2(k-\check{g})}  | \Omega \rangle \\
\  \\
J_{0} | \Omega \rangle =  \frac{1}{(k-\check{g})} S^{a}_{a0} | \Omega
\rangle = \frac{p}{(k-\check{g})}  | \Omega \rangle \\
\  \\
T_{0} |  a_{1}, \ldots, a_{n} \rangle = \frac{kN}{2(k-\check{g})}
|  a_{1}, \ldots, a_{n} \rangle \\
\  \\
J_{0} |  a_{1}, \ldots, a_{n} \rangle =  \frac{kN}{(k-\check{g})}
|  a_{1}, \ldots, a_{n} \rangle \\
\  \\
T_{0} | \bar{\Omega}, a_{1}, \ldots, a_{D} \rangle =
\frac{kD-\bar{p}}{2(k-\check{g})}| \bar{\Omega}, a_{1},
\ldots, a_{D} \rangle \\
\  \\
J_{0} | \bar{\Omega}, a_{1}, \ldots, a_{D} \rangle =
\frac{kD-\bar{p}}{2(k-\check{g})}| \bar{\Omega}, a_{1}, \ldots, a_{D} \rangle
\end{array}
\end{equation}
where the states $|\Omega \rangle$ and $|\bar{\Omega} \rangle$
belonging to the vacuum representation of $\hat{g}$ satisfy
\begin{equation}
\begin{array}{l}
U_{a0}  | \Omega \rangle =  0 \\
\  \\
U^{a}_{0}  | \bar{\Omega} \rangle =  0

\end{array}
\end{equation}
and
\begin{equation}
\begin{array}{l}
S^{a}_{a0}  | \Omega \rangle =  p  | \Omega \rangle \\
\  \\
S^{a}_{a0}  | \bar{\Omega} \rangle
=  -\bar{p}  |\bar{\Omega} \rangle
\end{array}
\end{equation}

The possible eigenvalues $p$ and $\bar{p}$ of $S^{a}_{a0}$ on the
states $| \Omega \rangle$ and $|\bar{\Omega} \rangle$ can be read off
from the results of \cite{oscmet} for $G=SU(m,n)$, $SO^{*}(2n)$,
 or
$Sp(2n,{\cal R})$. The same methods can be extended to the cases of
$SO(n,2)$, $E_{6}(-14)$, and $E_{7}(-25)$ so as to determine the
possible eigenvalues of $S^{a}_{a0}$ on the states $| \Omega \rangle$
and $|\bar{\Omega} \rangle$.  In \cite{oscmet}, the states
 $| \Omega \rangle$ ($|\bar{\Omega} \rangle$)
 that transform in an irreducible
representation of the maximal compact subgroup $H$ were called the
ground states of a highest (lowest) weight unitary irreducible
representation of $G$.  Thus the necessary requirement that only the
unitary representations of the non-compact group
 $G$ be admitted leads to an infinite set of
chiral (anti-chiral) primary states of the $N=2$ SCA.  However if we
further impose the condition
\begin{equation}
0 \leq h \leq \frac{c}{6}
\end{equation}
that follows from the unitarity of the $N=2$ module \cite{LVW} we are
restricted to a finite subset.
We should note that
the purely fermionic chiral (or anti-chiral) primary states
$| a_{1},\ldots,a_{N}\rangle$ ($| \bar{a}_{1},\ldots,\bar{a}_{N}\rangle$)
all satisfy the bound $ h \leq \frac{c}{6}$, just as in the compact case.

\section{The Chiral Primary States of $N=2$ Superconformal
 Algebras constructed over
Freudenthal Triple Systems}
\setcounter{equation}{0}
The exceptional Lie algebras $G_{2}, F_{4}$ and
$E_{8}$ do not admit a TKK type construction over a Jordan triple system
. A generalization
of the TKK construction to more general triple systems was given
by Kantor \cite{IK}. All finite dimensional simple Lie algebras
admit a realization over the Kantor triple systems (KTS). The Kantor's
  construction  of
Lie algebras was further developed and
 generalized to a unified construction of Lie
and Lie superalgebras in \cite{BG79}. This construction proceeds as follows.

Every simple Lie algebra $g$ admits  a 5-grading (Kantor structure) with
respect to some
subalgebra $g^{0}$ of maximal rank \cite{IK,BG79}:
\begin{equation}
g = g^{-2} \oplus g^{-1} \oplus g^{0} \oplus g^{+1} \oplus g^{+2}
\end{equation}

One associates  with the grade $+1$ subpace of $g$ a
triple system V and labels the elements  of $g^{+1}$ subspace with
the elements
of V \cite{IK,BG79}:
\begin{equation}
U_{a} \in g^{+1} \Longleftrightarrow a \in V
\end{equation}
Every simple Lie algebra $g$ admits a conjugation under which the grade
$+m$ subspace gets mapped into grade $-m$ subspace:

\begin{equation}
U^{a} \equiv U_{a}^{\dagger} \in g^{-1}
\Longleftrightarrow U_{a} \in g^{+1}
\end{equation}
One defines the commutators of $U_{a}$ and $U^{b}$ as
\begin{equation}
\begin{array}{l}
{[}U_{a} , U^{b}{]} = S_{a}^{b} \in g^{0}  \\
\ \\
{[}U_{a} , U_{b}{]} = K_{ab} \in g^{+2}               \\
\ \\
{[}U^{a} , U^{b}{]} = K^{ab} \in g^{-2}             \\
\ \\
{[}S_{a}^{b} , U_{c}{]} = U_{(abc)} \in g^{+1}

\end{array}
\end{equation}
where $(abc)$ denotes the triple (or ternary) product under which the
elements of V close. The remaining non-vanishing commutators of $g$ can
all be expressed in terms of the triple product $(abc)$.
The Jacobi identities of $g$  follow from  the following defining
identities of a KTS \cite{IK,BG79}
\begin{equation}
\begin{array}{l}

(ab(cdx)) - (cd(abx)) - (a(dcb)x) + ((cda)bx) = 0      \label{eq:K1}
\end{array}
\end{equation}
\begin{equation}
\begin{array}{l}
\{ (ax(cbd)) - ((cbd)xa) + (ab(cxd)) + (c(bax)d) \} - \{ c
\leftrightarrow d \} = 0                  \label{eq:K2}

\end{array}
\end{equation}
In general a given simple Lie algebra admits
several inequivalent such constructions
corresponding to different choices of the subalgebra $g^{0}$ and
different KTS's.

The construction of $N=2$ superconformal algebras over KTS's was given in
\cite{GH91,GH92}.
The realization of N=2 SCA's over the KTS's is equivalent
to their realization over the coset  spaces
G/H  where G and H are the groups generated by $g$ and $g^{0}$,
respectively.

One can extend the study of the chiral primary states of $N=2$ SCA's associated
with hermitian JTS's to their realization over the more general KTS's.
However , in this paper, we shall restrict our analysis to a
 very special subset of KTS's, namely the Freudenthal triple systems (FTS).
In \cite{GH91,GH92} it was shown that  the $N=2 $ SCA's
constructed over Freudenthal triple systems can be extended
to $N=4$ SCA's with the gauge group $SU(2) \times SU(2) \times U(1)$ .
The FTS's were first introduced by Freudenthal in his investigations of the
geometry of the
exceptional Lie groups
 \cite{HF} and studied in great detail
later \cite{TS,JF,KM,KS82}. There
 exists a one-
to-one correspondence between simple FTS's with a non-degenerate
bilinear form and finite dimensional simple Lie algebras \cite{KS82}.
For every simple Lie algebra $g$
 constructed over a Freudenthal triple system the
grade $\pm 2$ subspaces are one dimensional (except for $SU(2)$ for
which the grade $\pm 2$ subspaces vanish).
They generate an $SU(2)$ subalgebra of $g$ . Furthermore,
there is a universal relationship
between the dimension $D$ of a FTS and the dual coxeter number $
\check{g} $ of $g$ :
\begin{equation}
\check{g} = \frac{D}{2} + 2
\end{equation}
Every FTS admits a symplectic form $\Omega_{ab}$ such that
 the elements $K_{ab} (K^{ab})$ of the grade $+2(-2)$
subspaces can be represented as \cite{GH91,GH92} \\
\begin{equation}
\begin{array}{lll}
 K_{ab} & = & \Omega_{ab} K_{+}            \\
& &   \\
 K^{ab} & = & \Omega^{ab} K^{+}
\end{array}
\end{equation}
$\Omega^{ab}$ is the inverse of the symplectic form $\Omega_{ab}$
\begin{equation}
\begin{array}{l}
\Omega_{ab} \Omega^{bc} = \delta^{c}_{a} \\
\  \\
a,b,.. = 1,2,....,D
\end{array}
\end{equation}

For FT systems the expressions for the generators of N = 2 SCA take
the form \cite{GH91,GH92}
\begin{equation}
\begin{array}{lll}

G(z) & = & \sqrt{\frac{2}{k+ \check{g}}}
 \{ U_{a} \psi^{a} + K_{+} \psi^{+} -
\frac{1}{2} \Omega_{ab} \psi^{a} \psi^{b} \psi_{+} \}(z) \\
\end{array}
\end{equation}
\begin{equation}
\begin{array}{lll}
\bar{G}(z) & = & \sqrt{\frac{2}{k+ \check{g}}}
 \{ U^{a} \psi_{a} +
 K^{+} \psi_{+} -
 \frac{1}{2} \Omega^{ab} \psi_{a} \psi_{b} \psi^{+} \}(z)  \\
\end{array}
\end{equation}
\vspace{1.0cm}
\begin{equation}
\begin{array}{lll}
T(z) & = & \frac{1}{k+ \check{g}}
\{ \frac{1}{2}(U_{a}U^{a} + U^{a}U_{a})
+ \frac{1}{2}(K_{+}K^{+} + K^{+}K_{+})   \\
     &  &  \\
     &   & - \frac{k+1}{2}(\psi_{a} \partial \psi^{a} +
\psi^{a} \partial \psi_{a})
- \frac{1}{2}(k+ \check{g}-2)(\psi_{+} \partial \psi^{+} +
\psi^{+} \partial \psi_{+})   \\
     &  &  \\
      &   & + S_{a}^{b} \psi^{a} \psi_{b} +
\frac{1}{\check{g}-2} S_{a}^{a} \psi^{+} \psi_{+} +
       \psi_{+} \psi^{+} \psi^{a} \psi_{a} +
\frac{1}{4}\Omega_{ab} \psi^{a} \psi^{b} \Omega^{cd}
 \psi_{c} \psi_{d} \}(z)
\end{array}
\end{equation}
\vspace{1.0cm}
\begin{equation}
J(z) = \frac{1}{k+ \check{g}} \{ \frac{(\check{g} - 1)}{(\check{g}-2)}
S_{a}^{a} + (k+1)\psi^{a}
 \psi_{a} - (k- \check{g}+2) \psi^{+} \psi_{+} \}(z)
\end{equation}
where $ U_{a}(z),K_{+}(z) $ ,..etc represent the various graded subspaces
of the current algebra $\hat{g}$ .
The fermionic fields $\psi_{+} $ and $\psi^{+}$  are
associated with the grade $\pm2$ subspaces of $\hat{g}$ and satisfy
\begin{equation}
\psi^{+}(z) \psi_{+}(w) = \frac{1}{(z-w)} +...
\end{equation}

The central charge of the N = 2 SCA defined by
a FTS is
\begin{equation}
 c = \frac{6(k+1)(\check{g}-1)}{(k+ \check{g})}-3
\end{equation}
where $k$ is the level of $\hat{g}$ and $\check{g}$ is the dual coxeter
number. The above realization of $N=2$ SCA's over the simple FTS's correspond
 to the following coset spaces of simple Lie groups \cite{GH91,GH92}
\\
\begin{center}
\begin{tabular}{|c|c|} \hline
$G/H$ & $\check{g}$ \\
\hline
$SO(n)/SO(n-4)\times SU(2)\times U(1)$  & $n-2$  \\
$ SU(n)/SU(n-2)\times U(1)    $          & $n$    \\
$ Sp(2n)/Sp(2n-2)\times U(1)   $         & $n+1$  \\
$ G_{2}/SU(2)\times U(1)        $        & $4$    \\
$ F_{4}/Sp(6)\times U(1) $               & $9$    \\
$ E_{6}/SU(6)\times U(1)     $           & $12$   \\
$ E_{7}/SO(12)\times U(1)  $             & $18$   \\
$ E_{8}/E_{7}\times U(1)   $             & $30$  \\
\hline
\end{tabular}
\end{center}

The states $|\Omega \rangle $ belonging to the vacuum
 representation of $\hat{g}$
that are annihilated by
$U_{a0}$
\begin{equation}
U_{a0} |\Omega \rangle = 0
\end{equation}
are all chiral primary. (Note that the above condition implies also that
$K_{+0} |\Omega \rangle = 0 $). The corresponding $h$ and $q$ values are
\begin{equation}
h=\frac{1}{2} q = \frac{p(\check{g}-1)}{2(k+\check{g})(\check{g}-2)}
\end{equation}
where $p$ is the eigenvalue of $S_{a0}^{a}$ on the state $|\Omega \rangle $.
The corrresponding anti-chiral primary states $|\bar{\Omega} \rangle$
are those states belonging to a  highest weight representation $\hat{g}$ that
are annihilated by all the positive mode generators of $\hat{g}$ and that
satisfy
\begin{equation}
U^{a}_{0}|\bar{\Omega} \rangle = 0
\end{equation}
Their $h$ and $q$ values are
\begin{equation}
h=-\frac{1}{2}q = \frac{\bar{p}(\check{g}-1)}{2(k+\check{g})
(\check{g}-2)}
\end{equation}
where we denoted the eigenvalue of $S_{a0}^{a}$ on $|\bar{\Omega} \rangle $
as $-\bar{p}$.
Again the eigenvalues $p$ and $-\bar{p}$ of $S_{a0}^{a}$
 can be calculated rather simply for all the cases
by using the methods of references \cite{oscmet}.

 To discuss the purely fermionic
chiral primary states simply we choose a basis for the FTS such that
\begin{equation}
\begin{array}{l}
\Omega_{A,B} = 0   \hspace{1.0cm} A,B,.. =1,2,..., \frac{D}{2} \\
\  \\
\Omega_{A,B+\frac{D}{2}} = \delta_{A,B}  \\
\end{array}
\end{equation}
Then the states of the form
\begin{equation}
\begin{array}{l}
|A_{1},A_{2},...A_{N} \rangle \equiv \psi_{-\frac{1}{2}}^{A_{1}}
\psi_{-\frac{1}{2}}^{A_{2}}... \psi_{-\frac{1}{2}}^{A_{N}} |0\rangle
\\
\  \\
N = 0,1,2,..., \frac{D}{2}
\end{array}
\end{equation}
and the state
\begin{equation}
|\bar{s} \rangle = \psi_{+(-\frac{1}{2})} |0\rangle
\end{equation}
are all chiral primary and have the $q$ eigenvalues $\frac{(k+1)N}
{(k+\check{g})}$ and $\frac{(k-\check{g}+2)}{(k+\check{g})}$,
respectively.
Again the state $|0\rangle$ represents the tensor product of the Fock
vacuum with the scalar vacuum representation of $\hat{g}$.
The unique chiral primary state with $h=\frac{c}{6}$ is
\begin{equation}
|a_{1},....,a_{D},s \rangle \equiv \psi^{a_{1}}_{-\frac{1}{2}}
 \cdots \psi^{a_{D}}_{-\frac{1}{2}}
\psi^{+}_{-\frac{1}{2}} |0 \rangle
\end{equation}
Tensoring this state with the states
$|\bar{\Omega}\rangle$ belonging to the vacuum representation
of $\hat{g}$ that are annihilated by $U^{a}_{0}$ we obtain additional
chiral primary states
\begin{equation}
|\bar{\Omega},a_{1},...,a_{D},s \rangle
\end{equation}
with the $h$ values
\begin{equation}
h = \frac{c}{6} - \frac{\bar{p}(\check{g}-1)}{2(k+\check{g})(\check{g}-2)}
\end{equation}
The corresponding anti-chiral primary states are simply $
|\Omega, \bar{a}_{1},...,\bar{a}_{D},\bar{s} \rangle $.

We should note that the non-compact analogs of the coset spaces $G/H$
where $H$ is the maximal compact subgroup of $G$ associated with FTS's do not
exist as these spaces are not symmetric spaces.

\section{The Chiral and Anti-Chiral Rings in $N=4$ Superconformal Algebras}

The unitary representations of the maximal $N=4$ superconformal
algebras \cite{STV,STVS} were studied in \cite{GPTV}
 and their characters in
\cite{PT}. The maximal $N=4$ SCA has 4 supersymmetry generators, 4
dimension $\frac{1}{2}$  operators, and $SU(2) \times SU(2) \times
U(1)$ local gauge symmetry generators.  Our aim is to generalize the
concept of chiral (anti-chiral) rings to the maximal $N=4$ SCAs.
Following \cite{GPTV}, let us denote the supersymmetry generators as
$G^{+}$, $G^{-}$, $G^{+K}$, and $G^{-K}$,
the two $SU(2)$ currents as $A^{+i}$ and $A^{-i}$
 ($i=1,2,3$) , the $U(1)$
current as $W$ and the four dimension $\frac{1}{2}$ operators as
$Q^{+}$, $Q^{-}$, $Q^{+K}$, and $Q^{-K}$.  We shall work in the N-S
moding and assume the following hermiticity properties
\begin{equation}
\begin{array}{cc}
T_{n}^{\dagger} = T_{-n} &   \\
        \ & \ \\
(G^{+}_{s})^{\dagger} = G^{-}_{-s} ; & (G^{+K}_{s})^{\dagger} = G^{-K}_{-s} \\
  \  & \  \\
(A^{+i}_{n})^{\dagger} = A^{+i}_{-n}; & (A^{-i}_{m})^{\dagger} = A^{-i}_{-m} \\
\ & \  \\
W_{n}^{\dagger} = W_{-n} &    \\
                     \ & \ \\
(Q^{+}_{r})^{\dagger} = - Q^{-}_{-r}; & (Q^{+K}_{r})^{\dagger} = - Q^{-K}_{-r}

\end{array}
\end{equation}
where $m,n,\ldots = 0,\mp 1, \ldots$ and $r,s= \mp \frac{1}{2},
\mp \frac{3}{2}, \ldots$. \footnote{Note that $SU(2) \times SU(2) \times
U(1)$ current generators are hermitian.  In \cite{GPTV} they were taken as
anti-hermitian.}  The central charge of the $N=4$ SCA is given by
\begin{equation}
c=\frac{6k^{+}k^{-}}{k^{+}+k^{-}}=\frac{6k^{+}k^{-}}{k}
\end{equation}
where $k^{+}$ and $k^{-}$ are the levels of the two $SU(2)$ currrents and
$k=k^{+}+k^{-}$.  The highest weight state $|h,\ell^{+},\ell^{-},u \rangle$
of a unitary representation of the $N=4$ SCA is defined by the conditions
\begin{equation}
\begin{array}{l}
T_{n} |h,\ell^{+},\ell^{-},u \rangle
=  A^{\mp i}_{n} |h,\ell^{+},\ell^{-},u \rangle
              =  W_{n}|h,\ell^{+},\ell^{-},u \rangle = 0, \hskip 3ex n>0 \\
\   \\
G^{+}_{r} |h,\ell^{+},\ell^{-},u \rangle
 =  G^{-}_{r} |h,\ell^{+},\ell^{-},u \rangle = 0  \\
\  \\
 G^{+K}_{r}|h,\ell^{+},\ell^{-},u \rangle
 =  G^{-K}_{r} |h,\ell^{+},\ell^{-},u \rangle = 0,
 \hskip 3ex r>0 \\
             \  \\
Q^{+}_{r} |h,\ell^{+},\ell^{-},u \rangle
 =  Q^{-}_{r} |h,\ell^{+},\ell^{-},u \rangle = 0 \\
             \  \\
 Q^{+K}_{r}|h,\ell^{+},\ell^{-},u \rangle
 =  Q^{-K}_{r} |h,\ell^{+},\ell^{-},u \rangle = 0 ,

 \hskip 3ex  r>0 \\
          \   \\
A^{++}_{0}|h,\ell^{+},\ell^{-},u \rangle
 =  A^{-+}_{0}|h,\ell^{+},\ell^{-},u \rangle =0 \\
              \   \\
T_{0}|h,\ell^{+},\ell^{-},u \rangle  =  h |h,\ell^{+},\ell^{-},u \rangle  \\
  \  \\
(A^{+i}_{0}A^{+}_{0i})|h,\ell^{+},\ell^{-},u \rangle =  \ell^{+}(\ell^{+}+1)

|h,\ell^{+},\ell^{-},u \rangle  \\
   \ \\
(A^{-i}_{0}A^{-}_{0i})|h,\ell^{+},\ell^{-},u \rangle =  \ell^{-}(\ell^{-}+1)
|h,\ell^{+},\ell^{-},u \rangle  \\
  \  \\
W_{0}|h,\ell^{+},\ell^{-},u \rangle  = u |h,\ell^{+},\ell^{-},u \rangle

\end{array}
\end{equation}
where $A^{+ \mp}_{0} = A^{+1}_{0} \mp i A^{+2}_{0}$,
$A^{- \mp}_{0} = A^{-1}_{0} \mp i A^{-2}_{0}$.

The states belonging to the ``vacuum representation'' of the
$N=4$ SCA  for a given $h, \ell^{+}, \ell^{-}$ and $u$ will be denoted as

$|h,\ell^{+}_{3},\ell^{-}_{3},u \rangle$ which satisfy

\begin{equation}
\begin{array}{lll}
A^{+3}_{0}|h,\ell^{+}_{3},\ell^{-}_{3},u \rangle & = &
    \ell^{+}_{3} |h,\ell^{+}_{3},\ell^{-}_{3},u \rangle \\
          \ & \  \\
A^{-3}_{0}|h,\ell^{+}_{3},\ell^{-}_{3},u \rangle & = &
   \ell^{-}_{3} |h,\ell^{+}_{3},\ell^{-}_{3},u \rangle
\end{array}
\end{equation}
where $\ell^{+}_{3} = -\ell^{+}, -\ell^{+}+1, \ldots, \ell^{+}-1, \ell^{+}$
and     $\ell^{-}_{3} = -\ell^{-}, -\ell^{-}+1, \ldots, \ell^{-}-1, \ell^{-}$.

Any generalization of the concept of a chiral or anti-chiral primary ring
to $N=4$ SCAs must yield the standard definitions when truncated to the
$N=2$ subalgebra.  For example, the two supersymmetry generators
$G^{+}$ and $G^{-}$ generate an $N=2$ subalgebra
\begin{equation}
\{ G^{+}_{r}, G^{-}_{s} \} = L_{r+s} + \frac{1}{2}(r-s) J_{r+s} +
\frac{c}{6}(r^{2}-\frac{1}{4}) \delta_{r+s,0}
\end{equation}
where the $U(1)$ current $J(z)$ is given by
\begin{equation}
J(z) = \frac{2}{k} [k^{-}A^{+3}(z) + k^{+}A^{-3}(z)]
\end{equation}
When one looks for chiral primary states of the $N=2$ subalgebra among
the states belonging to the highest weight representation of the $N=4$
SCA, one is generally led to the trivial solution with $h=0$.
  This comes about
because the highest weight condition for the $N=4$ algebra requires
that the highest weight state be a Fock vacuum of
the four fermions $Q$.  Such a restriction has no counterpart for the
$N=2$ subalgebra.  In fact, one can show that the $N=2$ generators can
be decomposed as follows \cite{GPTV}

\begin{equation}
\begin{array}{l}
T(z)  =  \hat{T}(z) + T_{Q}(z) \\
     \   \\
G^{+}(z)   = \hat{G}^{+}(z) + G^{+}_{Q}(z) \\
                 \  \\
G^{-}(z)   =   \hat{G}^{-}(z) + G^{-}_{Q}(z) \\
                 \  \\
J(z)  =   \hat{J}(z) + J_{Q}(z)
\end{array}
\end{equation}
where the operator product of the hatted operators with those labelled
by $Q$ are regular.  The generators $T_{Q}$, $G^{\mp}_{Q}$, and
$J_{Q}$ are defined by a ``matter multiplet@ of the $N=2$ algebra
 \cite{GPTV}:
\begin{equation}
\begin{array}{l}
T_{Q}(z)  =  - \frac{1}{2}(A\bar{A} + \partial Q \bar{Q} + \partial \bar{Q}
Q)(z) \\
   \ \\
G^{+}_{Q}  =  - \frac{1}{\sqrt{2}}\bar{A}Q(z) \\
                        \  \\
G^{-}_{Q}  =  - \frac{1}{\sqrt{2}}A\bar{Q}(z) \\
                               \ \\
J_{Q}   =  - Q\bar{Q}(z)
\end{array}
\end{equation}
where
\begin{equation}
\begin{array}{l}
A  =  - \sqrt{\frac{2}{k}} (A^{+3}-A^{-3}+iW) \\
                        \  \\
\bar{A}  =  \sqrt{\frac{2}{k}} (A^{+3}-A^{-3}-iW) \\
                        \  \\
Q  =  \frac{2}{\sqrt{k}} Q^{+} \\
                  \  \\
\bar{Q}  =  \frac{2}{\sqrt{k}} Q^{-}
\end{array}
\end{equation}
Now if we denote the eigenvalues of $\hat{T}_{0}$, $T_{Q_{0}}$,
$\hat{J}_{0}$ and $J_{Q_{0}}$ as $\hat{h}$, $h_{Q}$, $\hat{q}$ and
$q_{Q}$, respectively we have from the results of \cite{LVW}
\begin{equation}
\begin{array}{ll}
\hat{h} & \geq \frac{|\hat{q}|}{2} \\
 &  \\
h_{Q} & \geq \frac{|q_{Q}|}{2}
\end{array}
\end{equation}
To have $h = \frac{|q|}{2}$ we must then have $\hat{h} =
\frac{|\hat{q}|}{2}$ and $h_{Q} = \frac{|q_{Q}|}{2}$.  On the highest
weight representation space of the $N = 4$ SCA we find that $q_{Q} =
0$, which requires that $h_{Q} = 0$ for the chiral primary states.
 Hence there is
no non-trivial contribution from the matter sector to the chiral
primary rings of the $N = 2$ subalgebra.  Therefore in looking for
chiral primary states we need only to restrict ourselves to the $N =
2$ SCA generated by $\hat{T}$, $\hat{G}$ and $\hat{J}$.

There is another truncation of the $N = 4$ SCA to an $N = 2$ subalgebra
 independent of the above one \cite{GPTV}.
  It is the $N = 2$  subalgebra generated by $G^{+K}$ and $G^{-K}$

\begin{equation}
\{G^{+K}_{r}, G^{+K}_{s}\} = T_{r+s} + \frac{1}{2}(r-s)J^{\prime}_{r+s} +
\frac{c}{6}(r^{2} - \frac{1}{4})\delta_{r+s,0}
\end{equation}
where
\begin{equation}
J^{\prime}(z) = \frac{2}{k}(k^{-}A^{+}_{3} - k^{+}A^{-}_{3})(z)
\end{equation}
Again the $N = 2$ generators can be decomposed as
\begin{equation}
\begin{array}{l}
T(z)  = \hat{T}^{\prime}(z) + T_{Q^{\prime}}(z)  \\
  \  \\
G^{+K}(z)  = \hat{G}^{+K}(z) + G^{+K}_{Q^{\prime}}(z)  \\
  \  \\
G^{-K}(z)  = \hat{G}^{-K}(z) + G^{-K}_{Q^{\prime}}(z)  \\
  \  \\
J^{\prime}(z)  = \hat{J}^{\prime}(z) + J_{Q^{\prime}}(z)
\end{array}
\end{equation}
In this case the ``matter multiplet" is
\begin{equation}
\begin{array}{l}
A^{\prime}(z)  = -\sqrt{\frac{2}{k}} (A^{+3} + A^{-3} + i W)(z)  \\
  \   \\
  \bar{A}^{\prime}(z)  = \sqrt{\frac{2}{k}} (A^{+3} + A^{-3} - i W)(z)  \\

  \  \\
Q^{\prime}(z)  = \frac{2}{\sqrt{k}}Q^{+K}(z)  \\
  \  \\
\bar{Q}^{\prime}(z)  = \frac{2}{\sqrt{k}}Q^{-K}(z)
\end{array}
\end{equation}
To find non-trivial solution for the conditions defining the chiral
primary states we need again to resrict ourselves to the hatted $N =
2$ subalgebra generated by $\hat{T}^{\prime}(z)$, $\hat{G}^{+K}(z)$,
$\hat{G}^{-K}(z)$ and $\hat{J}^{\prime}(z)$.

Since we have two independent $N = 2$ subalgebras of the $N = 4$ SCA (up
to automorphisms) the natural generalization of the concept of
 chirality or anti-chirality to the $N = 4$ case would be to look for
  states that are chiral or anti-chiral with respect to both $N = 2$
    subalgebras.  Thus we can define four rings in the $N = 4$ case namely
      the (cc), (aa), (ca), and (ac) rings \footnote{note these are not to
     be confused with the (c,c), (a,a), (c,a), and (a,c) rings of an $N = 2$
 SCA when are considers both left- and right-moving sectors}.  We shall
   refer to the subalgebras generated by $\{\hat{T}, \hat{G}, \hat{J}\}$
    and by $\{\hat{T}^{\prime}, \hat{G}^{\prime}, \hat{J}^{\prime}\}$ as
    the first and the second $N = 2$ superconformal algebra, repectively.
 The (cc) ring corresponds to the primary states $|\phi \rangle$
of the $N = 4$ SCA which  satisfy
\begin{equation}
G^{+}_{-\frac{1}{2}}|\phi \rangle =
G^{+K}_{-\frac{1}{2}}|\phi \rangle = 0
\end{equation}
The (aa) ring is defined by primary states $|\phi \rangle$
satisfying
\begin{equation}
G^{-}_{-\frac{1}{2}}|\phi \rangle =
G^{-K}_{-\frac{1}{2}} |\phi \rangle = 0
\end{equation}
The (ca) ring is defined by
\begin{equation}
G^{+}_{-\frac{1}{2}}|\phi \rangle =
G^{-K}_{-\frac{1}{2}} |\phi \rangle = 0
\end{equation}
and the (ac) ring by
\begin{equation}
G^{-}_{-\frac{1}{2}}|\phi \rangle =
G^{+K}_{-\frac{1}{2}} |\phi \rangle = 0
\end{equation}
Thus if we take both left and right-moving sectors into account we
have 16 different rings, eight of which are conjugates to the other
eight.
     We shall first give
an analysis of these rings within the ``massless
representations" of the $N = 4$ SCA's for which a complete treatment
was given in \cite{GPTV}.  Later we shall prove that massless
representations yield all possible rings of the $N = 4$ SCAs. In \cite
 {GPTV} massless representations were defined by the condition
\begin{equation}
\tilde{G}^{+}_{-\frac{1}{2}} |hws \rangle = 0
\end{equation}
where $|hws \rangle$ stands for a highest weight vector of the
 of the non-linear $N = 4$ algebra one obtains
by factoring out the four dimension $\frac{1}{2}$ operators and the
$U(1)$ current \cite{GPTV,GS,VK}. $\tilde{G}^{+}$ is one of the supersymmetry
generators of the nonlinear algebra.
Furthermore , for given levels $k^{+}$ and $k^{-}$ of the two $SU(2)$ currents
 algebras of the $N = 4$ SCA, the allowed angular momentum quantum
numbers $\ell^{+}$ and $\ell^{-}$ are
\begin{equation}
\begin{array}{ll}
\ell^{-} & = 0, \frac{1}{2},1,\dots,\frac{1}{2}(k^{-}-1)  \\
  &  \\
\ell^{+} & = 0, \frac{1}{2},1,\ldots,\frac{1}{2}(k^{+}-1)
\end{array}
\end{equation}

For $k^{-} = 1$ we have $\ell^{-} = 0$ and the restriction to an $N =
2$ subalgebra leads to the minimal discrete series representations of
the corresponding $N = 2$ subalgebra (with $\hat{c}<3$ or
$\hat{c}^{\prime}<3$). The allowed eigenvalues $\hat{q}$ of
$\hat{J}_{0}$ and $\hat{h}$ of $\hat{T}_{0}$ on the highest weight
representations are
\begin{equation}
\begin{array}{ll}
\hat{q} & = \frac{2}{(k^{+}+1)}\ell^{+}_{3} \\
  &  \\
\hat{h} & = \frac{1}{(k^{+}+1)} \{\ell^{+}(\ell^{+}+1) - (\ell^{+}_{3})^{2}\}
\end{array}
\end{equation}
Thus the chiral primary states of the first $N = 2$ subalgebra are the
states for which
\begin{equation}
\ell^{+}_{3} = \ell^{+}
\end{equation}
and anti-chiral states are those for which
\begin{equation}
\ell^{+}_{3} = -\ell^{+}
\end{equation}
where $\ell^{+} = 0,\frac{1}{2}, 1,\ldots,\frac{1}{2}(k^{+}-1)$.
For the second $N = 2$ subalgebra we find the same conditions since
\begin{equation}
\begin{array}{ll}
\hat{q}^{\prime} & = \frac{2}{(k^{+}+1)} \ell^{+}_{3}  \\
  &  \\
\hat{h}^{\prime} & = \frac{1}{(k^{+} + 1)} \{ \ell^{+}(\ell^{+} +1) -
(\ell^{+}_{3})^{2} \}
\end{array}
\end{equation}
for $k^{-} = 1$.  Therefore for $k^{-} = 1$ and $k^{+}$ arbitrary we
find that the (ca) and (ac) rings are trivial, consisting only of the
identity element corresponding to the vacuum state with $\ell^{+} =
\ell^{-} = h = u = 0$.  The (cc) ring has the form
\begin{equation}
(cc) = \{1,x,\ldots,x^{p-1}\}
\end{equation}
where $p = 2k^{+}-1$ and $x^{p} = 0$.  The (aa) ring is simply the
 conjugate of
the (cc) ring.
\begin{equation}
(aa) = \{ 1,\bar{x},\ldots,\bar{x}^{p-1} \}
\end{equation}
The other massless representation of the $N = 4$ SCA with $k^{-}>1$
can be obtained from the coset space \cite{GPTV}
\begin{equation}
\frac{SU(N)}{SU(N-2) \times SU(2) \times U(1)} \bigotimes SU(2) \times U(1)
\end{equation}
with the level of $SU(N)$ taken at $(k^{+}-1)$ and $k^{-} = N - 1$.
The eigenvalues of $\hat{T}_{0}$, $\hat{J}_{0}$, $\hat{T}^{\prime}$
and $\hat{J}^{\prime}$ on the highest weight representation space are
\begin{equation}
\begin{array}{ll}
\hat{h}  & = \frac{1}{k} \{(\ell^{+} - \ell^{-})^{2} -
(\ell^{+}_{3}-\ell^{-}_{3})^{2} + k^{-}\ell^{+} + k^{+}\ell^{-} \}   \\
  &  \\
\hat{q} & = \frac{2}{k} \{k^{-}\ell^{+}_{3} + k^{+}\ell^{-}_{3}\}    \\
  &  \\
\hat{h}^{\prime} & = \frac{1}{k} \{(\ell^{+} - \ell^{-})^{2}
-(\ell^{+}_{3}+\ell^{-}_{3})^{2} +k^{-}\ell^{+} + k^{+}\ell^{-} \}  \\
  &  \\
\hat{q}^{\prime} & = \frac{2}{k} \{k^{-}\ell^{+}_{3} - k^{+}\ell^{-}_{3}\}
\end{array}
\end{equation}
Therefore the states in the highest weight representation that are
chiral primary with respect to both $N = 2$ SCAs must satisfy
\begin{equation}
(cc) \Leftrightarrow \ell^{-} = 0 ,\; \ell^{+}_{3} = \ell^{+}
\end{equation}
Similarly the states that are anti-chiral with respect to both $N = 2$
SCAs satisfy
\begin{equation}
(aa) \Leftrightarrow \ell^{-} = 0 , \; \ell^{+}_{3} = -\ell^{+}
\end{equation}
The states that are chiral (anti-chiral) with respect to the first $N
= 2$ subalgebra and anti-chiral (chiral) with respect to the second
satisfy
\begin{equation}
\begin{array}{lll}
(ca)  \Leftrightarrow & \ell^{+} = 0 , & \ell^{-}_{3} = \ell^{-}  \\
  &   &  \\
(ac)  \Leftrightarrow & \ell^{+} = 0 , & \ell^{-}_{3} = -\ell^{-}
\end{array}
\end{equation}
Hence we have the following rings of $N = 4$ SCAs
\begin{equation}
\begin{array}{ll}
(cc) & = \{1,x,\ldots,x^{p-1}\}  \\
  &  \\
(aa) & = \{1,\bar{x},\ldots,\bar{x}^{p-1}\}  \\
  &  \\
(ca) & = \{1,y,\ldots,y^{q-1}\}  \\
  &  \\
(ac) & = \{1,\bar{y},\ldots,\bar{y}^{q-1}\}
\end{array}
\end{equation}
with the restrictions
\begin{equation}
\begin{array}{l}
 x^{p} = 0 \\
   \\
 y^{q} = 0
\end{array}
\end{equation}
where $p = 2k^{+}-1$ and $q = 2k^{-}-1$.

Let us now prove that all the rings associated with $N = 4$ SCAs are
of the above form.  Consider for example a general (cc) ring.  Any
primary state belonging to the (cc) ring is annihilated by the
operators $\hat{G}^{+}_{-\frac{1}{2}}$, $\hat{G}^{K+}_{-\frac{1}{2}}$,
$\hat{G}^{-}_{\frac{1}{2}}$, and $\hat{G}^{-K}_{\frac{1}{2}}$ and hence by
 their anti-commutators
\begin{equation}
\begin{array}{ll}
\{\hat{G}^{+}_{-\frac{1}{2}}, \hat{G}^{-}_{\frac{1}{2}}\} & = \hat{T}_{0} -
\frac{1}{k}(k^{-}\hat{A}^{+3}_{0} + k^{+}\hat{A}^{-3}_{0}) \\
  &  \\
\{\hat{G}^{+K}_{-\frac{1}{2}},
\hat{G}^{-K}_{\frac{1}{2}}\} & = \hat{T}^{\prime}_{0} -
\frac{1}{k}(k^{-}\hat{A}^{+3}_{0} - k^{+}\hat{A}^{-3}_{0}) \\
  &  \\
\{\hat{G}^{+}_{\frac{1}{2}}, \hat{G}^{+K}_{-\frac{1}{2}}\} & =
- \frac{k^{-}}{k}\hat{A}^{++}_{0}  \\
  &  \\
\{\hat{G}^{+}_{-\frac{1}{2}}, \hat{G}^{+K}_{\frac{1}{2}}\} & =
\frac{k^{-}}{k}\hat{A}^{++}_{0}  \\
  &  \\
\{\hat{G}^{+}_{-\frac{1}{2}}, \hat{G}^{-K}_{\frac{1}{2}}\} & =
\frac{k^{+}}{k}\hat{A}^{-+}_{0}  \\
  &  \\
\{\hat{G}^{-}_{\frac{1}{2}}, \hat{G}^{+K}_{-\frac{1}{2}}\} & =
\frac{k^{+}}{k}\hat{A}^{--}_{0}
\end{array}
\end{equation}
where the hatted currents $\hat{SU}(2)^{+}$ and $\hat{SU}(2)^{-}$ have
levels $(k^{+}-1)$ and $(k^{-}-1)$, respectively. It is obvious from
the above that any state belonging to the (cc) ring must be an
$SU(2)^{-}$ singlet and satisfy
\begin{equation}
\begin{array}{l}
\ell^{+}_{3} = \ell^{+} , \ell^{-} = 0 \\
  \\
\hat{h} = \frac{k^{-}}{k}\ell^{+} = \frac{\hat{q}}{2} \\
   \\
\hat{h} = \hat{h}^{\prime} , \hat{q} = \hat{q}^{\prime}
\end{array}
\end{equation}
Similarly we find that any state belonging to the (aa) ring must be
annilhilated by $\hat{G}^{-}_{-\frac{1}{2}}$,
$\hat{G}^{-K}_{-\frac{1}{2}}$, $\hat{G}^{+}_{\frac{1}{2}}$,
$\hat{G}^{+K}_{\frac{1}{2}}$ and their anti-commutators, which leads
to the conditions
\begin{equation}
\begin{array}{l}
\ell^{+}_{3} = -\ell^{+} ; \; \ell^{-} = 0 \\
  \\
\hat{h} = \frac{k^{-}}{k}\ell^{+} = - \frac{\hat{q}}{2} \\
  \\
\hat{h} = \hat{h}^{\prime} ; \; \hat{q} = \hat{q}^{\prime}
\end{array}
\end{equation}

For the (ca) ring we find that the corresponding primary states must
satisfy
\begin{equation}
\begin{array}{l}
\ell^{-}_{3} = \ell^{-} ; \; \ell^{+} = 0 \\
  \\
\hat{h} = \frac{k^{+}}{k}\ell^{-} = \frac{\hat{q}}{2} \\
  \\
\hat{h} = \hat{h}^{\prime} , \; \hat{q} = -\hat{q}^{\prime}
\end{array}
\end{equation}
and for the (ac) ring
\begin{equation}
\begin{array}{l}
\ell^{-}_{3} = -\ell^{-} ; \;  \ell^{+} = 0 \\
  \\
\hat{h} = \frac{k^{+}}{k}\ell^{-} = -\frac{\hat{q}}{2} \\
  \\
\hat{h} = \hat{h}^{\prime} , \; \hat{q} = -\hat{q}^{\prime}
\end{array}
\end{equation}
The allowed angular momentum quantum numbers $\ell^{+}$ and $\ell^{-}$
are
\begin{equation}
\begin{array}{ll}
\ell^{+} & = 0,\frac{1}{2},\ldots,\frac{1}{2}(k^{+} -1) \\
  &  \\
\ell^{-} & = 0,\frac{1}{2},\ldots,\frac{1}{2}(k^{-} -1)
\end{array}
\end{equation}
Therefore the rings we obtained from the massless unitary highest
weight representations of the $N = 4$ SCA exhausts all the allowed
values of the quantum numbers and is complete.

{\it Acknowledgements:} Useful discussions with Bong Lian, Cumrun Vafa,
 Nick Warner
and Gregg Zuckerman are gratefully acknowledged.


\newpage
\begin{thebibliography}{99}
\bibitem{GSW} M. Green, J. H. Schwarz and E. Witten, "Superstring
Theory" , Cambridge Univ. Press, (1987, Cambridge).

\bibitem{IS} See P. Fendley, W. Lerche, S. D. Mathur and N.P. Warner,
"N=2 Supersymmetric Integrable Models from Affine Toda Theories",
Preprint CTP1865 (CALT-68-1631; HUTP-90/A036) and the references
therein.

\bibitem{TFT}  E. Witten,
{\it Nucl. Phys.} {\bf B340} (1990) 281 ; E. Witten,
{\it Comm. Math. Phys.} {\bf 118} (1988) 411 ;
C.  Vafa, Harvard Preprint HUTP-90/A064 ;
 J. Distler and P. Nelson,
{\it Phys. Rev. Lett.}
{\bf 66} (1991) 1955 ; R. Dijkgraaf, H. Verlinde and E.
Verlinde, Preprint PUPT-1217 (IASSNS-HEP-90/80 ).

\bibitem{BDFM} T. Banks, L. Dixon, D. Friedan and E. Martinec,
{\it Nucl. Phys.} {\bf B299} (1988) 613.
\bibitem{N=2} W. Boucher, D. Friedan and A. Kent,
{\it Phys. Lett.} {\bf 172B }
(1986) 316 ; P. DiVecchia, J.L. Petersen and M. Yu,
{\it Phys. Lett.}
{\bf 172B} (1986) 211 ; A.B. Zamolodchikov and V.A. Fateev,
{\it Zh. Eksp.
Theor.
Fiz.} {\bf 90} (1986) 1553 and {\it Sov. Phys.}
{\it JETP} {\bf 6} (1985) 215 ; Z. Qiu,
{\it Phys. Lett.}
{\bf 188B} (1987) 207; D. Gepner and Z. Qiu,
{\it Nucl. Phys.} {\bf B285} (1987) 423 ;

\bibitem{LVW} W. Lerche, C. Vafa and N.P. Warner,
{\it Nucl. Phys.} {\bf B324} (1989) 427.


\bibitem{KS89}Y. Kazama and H. Suzuki,
{\it Nucl. Phys.} {bf B321} (1989) 232;
{\it Phys. Lett.} {\bf B216} (1989) 112;


\bibitem{DLP} L.J. Dixon, J. Lykken and
 M.E. Peskin, {\it Nucl. Phys.} {\bf B325} (1989) 329.
\bibitem{IB} I. Bars,
{\it Nucl. Phys.} {\bf B334} (1990) 125;

\bibitem{GKO} P. Goddard, A. Kent and D. Olive,
{\it Phys. Lett.} {\bf 152B}
(1985) 88 ;
{\it Comm. Math. Phys.} {\bf 103} (1986)105 ; P. Goddard, W. Nahm
and
D. Olive, {\it Phys. Lett.} {\bf 160B} (1985) 111.

\bibitem{NS}N. Seiberg,{\it Nucl. Phys.} {\bf B303} (1988) 286.

\bibitem{mirror} see ,for example, C. Vafa, ``Topological Mirrors and
Quantum Rings'', Harvard Preprint HUTP-91/A059 and B. R. Greene and
M. R. Plesser, ``Mirror Manifolds: A Brief Review and Progress Report'',
Preprint CLNS91-1109 (YCTP-P32-91) and the references therein.

\bibitem{EY} T. Eguchi and S. -K. Yang,
{\it Mod. Phys. Lett.} {\bf A5} (1990) 1693;

\bibitem{MG91} M. G\"{u}naydin,
{\it Phys. Lett.} {\bf B255} (1991) 46.
\bibitem{GH91} M. G\"{u}naydin and S. Hyun, {\it Mod. Phys. Lett.}
 {\bf A6}
(1991) 1733.
\bibitem{GH92} M. G\"{u}naydin and S. Hyun, {\it Nucl. Phys.} {\bf B373}
(1992) 688.
\bibitem{TKK} J. Tits,
{\it Nederl. Akad. van Wetens.} {\bf 65} (1962) 530 ;
I. L. Kantor, {\it Sov. Math. Dok.} {\bf 5} (1964) 1404 ; M. Koecher,
{\it Amer. J. Math.}
{\bf 89} (1967) 787.

\bibitem{BBSS} F. A. Bais, P. Bouwknegt, K. Schoutens and M. Surridge
, {\it Nucl. Phys.} {\bf B304} (1988)348.

\bibitem{OL} See O. Loos,{\it Bounded Symmetric Domains and Jordan
Pairs} (Mathematical Lectures, Univ. of California, 1977) and the
references therein.

\bibitem{LZ} B. H. Lian and G.
J. Zuckerman,
{\it Comm. Math. Phys.} {\bf 135} (1991) 547;

\bibitem{GNO} P. Goddard, W. Nahm and D. Olive, {\it Phys. Lett.}
{\bf 160B} (1985) 111.

\bibitem{KN} S.Kobayashi and K. Nomizu, `` Foundations of Differential
Geometry'' , Interscience Publishers, 1963.

\bibitem{HC} Harish-Chandra, {\it Am. J. Math.} {\bf 77} (1955) 743: {\bf 78}
(1956) 1.

\bibitem{oscmet} M. G\"{u}naydin and C. Saclioglu,{\it Comm. Math. Phys.}
{\bf 87} (1982) 159;
 I. Bars and M. G\"{u}naydin,{\it Comm. Math. Phys.}{\bf 91} (1983)
31 ; M. G\"{u}naydin, P. van Nieuwenhuizen and N. P. Warner,
{\it Nucl. Phys.}{\bf B255} (1985) 63 ; M. G\"{u}naydin and N. P. Warner,
{\it Nucl. Phys.}{\bf B272} (1986) 99 ; M. G\"{u}naydin and N. Marcus,
{\it Class. Quant. Grav.}{\bf 2} (1985) L11 ; M. G\"{u}naydin, G. Sierra
and P. K. Townsend, {\it Nucl. Phys.}{\bf B274} (1986) 429 ; M. G\"{u}naydin,
{\it J. Math. Phys.}{\it 29} (1988) 1275 ;
 M. G\"{u}naydin and S. Hyun, {\it J. Math.
Phys.} {\bf 29}(1988) 2367 ; M. G\"{u}naydin and R. Scalise, {\it J. Math.
Phys.} {\bf 32} (1991) 599.


\bibitem{IK} I.L. Kantor,
{\it Trudy Sem. Vektor. Anal.} {\bf 16} (1972) 407 ;
{\it Sov. Math. Dokl.} {\bf 14} (1973) 254.
\bibitem{BG79} I. Bars and M. G\"{u}naydin, {\it J. Math. Phys.} {\bf 20}
(1979)1977.
\bibitem{HF} H. Freudenthal,
{\it Proc. Konikl.Nederl. Akad. Wet. Ser.} {\bf A57}
(1954) 218-230 and 363-408 ; K. Meyberg, ibid
{\bf A71} (1968) 162-174 and
175-190.
\bibitem{TS} T. A. Springer, {\it Indag. Math.} {\bf 24} (1962) 259 .

\bibitem{JF} J. C. Ferrar,
{\it Trans. Amer. Math. Soc.} {\bf 155} (1971) 397.

\bibitem{KM} K. Meyberg, {\it Proc. Kon. Ned. Acad. Wetens. SerA}
 {\bf 71}
(1968) 162.
\bibitem{KS82}I.L. Kantor and I.M. Skopets,{\it Sel. Math. Sov.} {\bf 2}
(1982)293.
\bibitem{STV}K. Schoutens,
{\it Nucl. Phys.} {\bf B295} (1988) 634 ; A. Sevrin, W.
Troost and A. Van Proeyen,
{\it Phys. Lett.} {\bf B208} (1988) 447.
\bibitem{STVS}A. Sevrin, W. Troost, A. Van Proeyen and Ph. Spindel,
{\it Nucl. Phys.} {\bf B311} (1988) 465 ; A. Van Proeyen,
{\it Class. Quantum Gravity} {\bf 6}
(1989) 1501
; A.Sevrin and G. Theodoridis, {\it Nucl. Phys.} {\bf B332} (1990) 380.
\bibitem{GPTV} M. G\"{u}naydin, J.L. Petersen, A. Taormina and A. Van
Proeyen, {\it Nucl. Phys.} {\bf B322} (1989)402.
\bibitem{PT} J. L. Petersen and A. Taormina, {\it Nucl. Phys. }
{\bf B333} (1990)833 ; {\bf B354} (1991) 689; H. Ooguri, J. L. Petersen and
A. Taormina, Preprint EFI-91-15 (May 1991).
\bibitem{GS} P. Goddard and A. Schwimmer, {\it Phys. Lett.} {\bf B214}
(1988) 209.
\bibitem{VK} V.G. Knizhnik, {\it Theor. Math. Phys.} {\bf 66} (1986) 68.



\end{thebibliography}
\end{document}